\documentclass[reprint,superscriptaddress,nofootinbib,amsmath,amssymb, aps]{revtex4-1}

\usepackage[a4paper, margin=2.50cm]{geometry}

\usepackage{bbm}
\usepackage{graphicx}
\usepackage{fancyhdr, color}
\usepackage{hyperref} 
\usepackage[normalem]{ulem}

\newcommand{\Time}{\mathcal{T}}

\newcommand{\average}[1]{\langle #1 \rangle}
\newcommand{\Trace}{{\rm Tr}}

\def \ket#1{\mathinner{|{#1}\rangle}}
\def \bra#1{\mathinner{\langle{#1}|}}

\def\braket#1{\mathinner{\langle{#1}\rangle}}

\newcommand{\ketbra}[2]{{\mathinner{| {#1} \rangle \langle {#2} |}} }

\newcommand{\matrixel}[3]{{\mathinner{\langle{#1}| {#2} | {#3}\rangle}} }

\newcommand{\Forder}{\overrightarrow{T}}

\newcommand{\GFunc}{\mathcal{G}}
\newcommand{\Prob}{\mathcal{P}}

\def \tmp{TMP}

\newcommand{\uba}{Departamento de F\'isica, Facultad de Ciencias Exactas y Naturales, Universidad de Buenos Aires and IFIBA, CONICET, Ciudad Universitaria, 1428 Buenos Aires, Argentina}

\newcommand{\chwqmbs}{{\color[rgb]{0.4,0.2,0.9} \sc Qbook:Ch.11}}

\begin{document}

\fancyhead[C]{\sc \color[rgb]{0.4,0.2,0.9}{Quantum Thermodynamics book}}
\fancyhead[R]{}
\fancyhead[L]{}

\title{Ancilla-assisted measurement of quantum work}

\author{Gabriele De Chiara}
\email{g.dechiara@qub.ac.uk} 
\affiliation{Centre for Theoretical Atomic, Molecular and Optical Physics, School of Mathematics and Physics, Queen's University, Belfast
BT7 1NN, United Kingdom}

\author{Paolo Solinas}
\email{paolo.solinas@spin.cnr.it} 
\affiliation{SPIN-CNR, Via Dodecaneso 33, 16146 Genova, Italy}

\author{Federico Cerisola} \affiliation{\uba}
\author{Augusto J. Roncaglia} \affiliation{\uba} 

\date{\today}

\begin{abstract}
We review the use of an external auxiliary detector for measuring the full distribution of the work performed on or extracted from a quantum system during a unitary thermodynamic process. We first illustrate two paradigmatic schemes that allow one to measure the work distribution: a Ramsey technique to measure the characteristic function and a positive operator valued measure (POVM) scheme to directly measure the work probability distribution. Then, we show that these two ideas can be understood in a unified framework
for assessing work fluctuations  through a generic quantum detector and describe two  protocols that are able to yield complementary information. This allows us also to highlight how quantum work is affected by the presence of coherences in the system's initial state.
Finally, we describe physical implementations and experimental realisations of the first two schemes.
\end{abstract}

\maketitle

\thispagestyle{fancy}

\section{Introduction} \label{sec:intro}

The concept of work is one of the cornerstones of classical physics.
In classical mechanics, it is defined as the integral of the force applied by an external agent times the displacement of the system.
It is deeply related to the concept of heat through the principle of energy conservation.
Heat is the dissipated energy due to the presence of non-conservative forces acting on the system.
Work and heat allow us to answer to a fundamental and practical question: how much energy does a system need to perform a specific task?
The advantages and elegance of such quantities relies on the fact that they can be calculated neglecting the details of the system and the dynamics.
    
Surprisingly, the discussion on how to extend the concept of work to the quantum domain has not been faced until recently and was motivated by a renewed interest in the study of non-equilibrium quantum systems, and the extension of classical fluctuation theorems \cite{Jarzynski1997Nonequilibrium,Crooks1999} to the quantum domain~\cite{kurchan2000quantum, tasaki2000jarzynski, campisi2011colloquium, esposito2009nonequilibrium}. Classical fluctuation theorems are powerful tools that go beyond the linear response theory, relating for instance the work statistics with equilibrium quantities such as free energy differences. 

The current  interest in out-of-equilibrium quantum thermodynamics has been triggered by recent advances in the coherent manipulation of elementary quantum systems. 
However, as discussed in other chapters of this book, the definition of work in a quantum setting is rather problematic. Quantum mechanics is built around the operators which are associated to  physical observables \cite{dirac1967principles}.
If we perform a {\it projective} measurement of an observable on a system, the latter collapses onto one of the eigenstates of the corresponding operator with a certain probability \cite{dirac1967principles}.
From this simple rule, we know how to calculate the statistics of the observable.

In this framework, the measurement of the system leads to two important implications. 
First, the measurement process is local in time, i.e. it is assumed to occur on timescales smaller than any other time scale and the wave-function collapse is supposed to be instantaneous. This also implies that the information about the observable and the associated operators are local in time.
Second, the measurement destroys any superposition of eigenstates of the measured operator  and, therefore, strongly perturbs the quantum system.

These two points caused some debates in the attempt to unambiguously define quantum work. 
To understand this point, let us consider a closed system in which all the energy injected by an external field is transformed in internal energy of the system, i.e. there is no dissipation or heat.
For a classical system, if the initial and final energies are $E_i$ and $E_f$, respectively, then the work performed is $W=E_f - E_i$. Therefore, $W$ depends on the values of the observable energy at two different times and it is thus non-local in time.

As discussed above,  observables in quantum mechanics are usually associated to Hermitian operators that are local in time. If we denote with $H_f $ and $H_i$ the final and initial Hamiltonian operators of the system, respectively, a plausible definition of a work operator could be $W = H_f - H_i$ \cite{engel2007jarzynski,allahverdyan2014}.
Though this definition is formally possible and is sometimes useful to connect the work fluctuations with macroscopic measurable quantities~\cite{Fusco, Villa}, it has no meaning in terms of quantum observables and measurements. 
Work is defined as the energy difference after a given process, and  as such it characterises a process and not the system's state. In addition, the number of possible values of work typically exceeds the dimension of the system, thus work cannot be represented as a Hermitian operator.
This point has generated confusions for a long time \cite{engel2007jarzynski,allahverdyan2014} and it has been clarified in Ref. \cite{talkner2007fluctuation}. 

We arrive to the following important observation:
Since quantum work does not meet the requirements to be associated to a standard (local in time) operator, the only way to clarify its meaning is to give an operative definition, that is, to describe the scheme we use to measure it.

The first proposal in this direction was done in some early papers \cite{kurchan2000quantum, tasaki2000jarzynski, campisi2011colloquium, esposito2009nonequilibrium} and it is described in chapter \chwqmbs   
of this book.
It is based on a double and sequential measurement of the system and we refer to it as the ``Two Measurement Protocol" (\tmp{}). The scheme is as follows:
The system energy $H_i$ is measured at the beginning of the evolution, then the system evolves under the effect of the driving field and the final energy $H_f$ is measured again at the end of the evolution.
The difference of the two energy measurements gives us information about the work statistics \cite{kurchan2000quantum, tasaki2000jarzynski,talkner2007fluctuation}.
This approach was initially proposed in the context of quantum fluctuation theorems and quantum Jarzynski equalities \cite{campisi2011colloquium, esposito2009nonequilibrium}.
It was later extended to describe the dissipated heat in open quantum systems in the weak and strong coupling regime \cite{campisi2011colloquium, campisi2009fluctuation, Carrega2016Energy}.
An advantage of this definition is that it clearly explains the meaning and the statistics of quantum work.

It is worth mentioning at this stage that, although quantum work is not a quantum observable \cite{talkner2007fluctuation}, i.e. it is not the expectation value of a Hermitian operator, it can be formally described as a generalised measurement. Quantum work can in fact be described as a positive operator valued measure (POVM)~\cite{roncaglia2014work} which is very common in quantum optics and information. This was an important conceptual result because it clarified the role of work in quantum mechanics. A POVM can always be interpreted (or realised) as a projective measurement in a larger Hilbert space.

If we consider a generic initial state, i.e. not an equilibrium thermalised state, the \tmp{} is unsatisfactory and incomplete because of the collapse of the quantum state induced by the initial measurement.
If the system is in a coherent superposition of eigenstates of $H_i$, the initial measurement destroys the quantum coherences and it completely changes the system dynamics.
This has both fundamental and practical implications.
From a fundamental point of view, we would like to understand if there are more general and less invasive measurement protocols that allow us to preserve the full quantum dynamics. 
From a practical point of view, the \tmp{} shows an important limitation since it is not possible to answer to the question: what is the energy needed to run a quantum device?

To weigh properly the implications of the last point, we discuss a specific example.
Suppose that we have a quantum computer that  runs the Grover quantum database search \cite{nielsen-chuang_book, Grover1996}.
If the database is composed of $N$ objects and the basis of all the possible logical states is denoted with $\{ \ket{x} \}$  ($x=0, 1, ..., N-1$),  the algorithm is able to find a particular marked string $\ket{\bar{x}}$.
As usual, let us suppose that the logical states are encoded in $n$ qubits ($N=2^{n}$) so that they can be written as logical strings of $0$ and $1$, e.g., $\ket{0} = \ket{00....0}$, $\ket{1} = \ket{10....0}$ and so on \cite{nielsen-chuang_book}.
The algorithm starts from the equal superposition state $\ket{\psi} = 1/\sqrt{N} \sum_x \ket{x}$ and, by means of a unitary evolution $U_{\rm algo}$, transforms it in the solution state  $\ket{\bar{x}}$, i.e., 
$\ket{\psi} \rightarrow U_{\rm algo} \ket{\psi} = \ket{\bar{x}}$.
We want to know what is the work necessary to run the quantum database search algorithm.

To answer to this question, we must first describe the energy spectrum of a single qubit.
Let a single qubit have energy $\epsilon$ and $-\epsilon$ for the eigenstates $\ket{0}$ and $\ket{1}$. 
Being a string of $\ket{0}$ and $\ket{1}$, every logical state $ \ket{x}$ has a defined energy that could be measured projectively.
However, the initial state $\ket{\psi}$ is not an eigenstate of the Hamiltonian and therefore it has no definite energy.
Using the \tmp{} to determine the work would lead to a collapse of $\ket{\psi}$ into a random state $ \ket{x}$.
The dynamics induced by $U_{\rm algo}$ would not give the solution $\ket{\bar{x}}$ and the final measurement would give another random value for the final energy.
Therefore, the \tmp{} makes it impossible to run the Grover algorithm because it changes completely the dynamics and, at the same time, is unable to determine what is the energy needed to run it.
Furthermore, for an initial state with coherences in the energy eigenbasis, the mean value of work  
obtained from the \tmp{} probability distribution and the difference between the mean initial energy and the mean final energy are in general different.
Indeed in Ref.~\cite{Perarnau-Llobet2017}, an important result has been put forward: there is no scheme able to estimate work in closed systems that produces outputs whose mean value is given by the difference in mean energies and coincides with the \tmp{} for initial states without coherences (see also Ref.~\cite{HayashiPRA2017}). This result states that the statistics of work for initial states with coherences cannot be defined via a probability distribution compatible with the above conditions. However, as we shall see in this chapter, it is possible to define a quasi-probability distribution which can assume negative values, signalling initial quantum coherence.

These examples point out the need for a more general operational definition of work.
One way to do this is to include an ancilla or quantum detector and to take into consideration its interaction with the quantum system under scrutiny and the perturbation the latter induces in the former.

Already several works suggested the use of ancillas for extracting the statistics of work. In Refs.~\cite{dorner2013extracting,mazzola2013measuring} a Ramsey scheme using an auxiliary qubit was proposed to measure the work characteristic function. Its experimental realisation with nuclear magnetic resonance was reported in Ref.~\cite{Batalho2014}. In Ref.~\cite{roncaglia2014work} a different approach was taken in which a detector, for example the position of a quantum particle or a light mode \cite{DeChiaraNJP2015}, is coupled to the system to extract directly the work probability distribution after a single quantum measurement. The scheme was recently realised with cold atoms \cite{Cerisola2017}.

These two schemes, reviewed in Sec.~\ref{sec:examples}, can be understood in a more general framework:  the interaction between the evolving system and the detector allows us to encode the information about the work performed on the system in the quantum state of the detector.
We show below that there are several ways, that we call {\it protocols}, to extract this information.
The choice between them depends on what kind of information we are interested in and if we want to preserve or not some of the quantum features of the dynamics.
Therefore, in this general setting, the quantum work statistics can change depending on the measurement protocol used.

A measurement protocol similar to the one proposed in the full counting statistics approach \cite{levitov1993charge, levitov1996electron, nazarov2003full, Belzig2001, Belzig2003} is able to completely preserve the coherence effect in the system.
It leads to a quasi-probability of work distribution in which negative probability regions are signatures of the quantumness of the system \cite{solinas2015fulldistribution,solinas2016probing, solinas2017measurement, clerk2011full, hofer2016,Hofer2017quasiprobability}.

Alternatively, with a measurement protocol conceptually similar to the von Neumann measurement scheme \cite{VonNeumann1955}, we directly measure the work distribution that in the limit of a precise measurement coincides with the \tmp{} scheme.
In this case, the uncertainty in the measurement plays a fundamental role since imprecise measurements lead to a revival of quantum dynamics and coherence effects.

These protocols can be unified in a single approach where the only difference is the final measurement of two non-commuting observables, e.g. momentum and position of a particle detector.
All the main protocols proposed in the literature can be revisited and framed in this more general measurement scheme that we explain in detail in Sec.~\ref{sec:general_framework}. In Sec.~\ref{sec:physical}, we discuss physical implementations and experimental realisations of the schemes. Finally, in Sec.~\ref{sec:conclusions}, we summarise and conclude.

\section{Measuring the work distribution with ancillas}
\label{sec:examples}

\subsection{Preliminaries on the work probability distribution in quantum mechanics}

One of the most common definitions of work is the \tmp{} scheme. We assume  a quantum system to be initially prepared in the state $\rho_S(0)$ and subject to the Hamiltonian $H_S(0)$. The initial energy of the system is measured yielding the energy eigenvalue $\epsilon_i^0$ while the system collapses to the state $\ket{\epsilon_i^0}$. The system's Hamiltonian is then changed in time inducing an evolution until the final time $\Time$ described by the operator $U_S(\Time)$. The final energy, described by the final Hamiltonian $H_S(\Time)$, is measured again yielding the result $\epsilon_j^\Time$. For this particular quantum trajectory the work performed on or extracted from the system is $W=\epsilon_j^\Time - \epsilon_i^0$, i.e. the difference of the final and initial energy. As the results of the measurements are stochastic with probabilities dictated by quantum theory, we find the probability distribution of work by summing over all possible measurement outcomes:
\begin{equation}
\label{eq:Pwtmp}
 \Prob(W)
 = \sum_{i} P_i \sum_j P_{i\rightarrow j} \delta(W-\epsilon_{j}^\Time + \epsilon_{i}^0) 
\end{equation}
where $P_i = \matrixel{\epsilon_{i}^0}{ \rho_S(0)}{\epsilon_{i}^0}$ is the probability that the first energy measurement results in $\epsilon_i^0$ and $P_{i\rightarrow j} = |\matrixel{\epsilon_{j}^\Time}{ U(\Time) }{\epsilon_{i}^0}|^2$ is the conditional probability of obtaining $\epsilon_j^\Time$ in the last measurement given that the result of the first measurement was $\epsilon_i^0$.

An equivalent description of the work statistics is to consider the Fourier transform of the work probability distribution known as the characteristic function of work:
\begin{equation}
\chi_\lambda = \int dW e^{i\lambda W} \Prob(W)
\end{equation}
which can be cast in the form of two-time correlation function~\cite{talkner2007fluctuation,campisi2011colloquium}:
 \begin{equation}
\label{eq:Gramsey}
\chi_\lambda = {\rm Tr}_S \Big[   U_S^\dagger(\Time) e^{i\lambda H_S(\Time)}U_S(\Time) e^{-i\lambda H_S(0)}\tilde\rho_S(0)   \Big].
\end{equation}
where $\tilde\rho_S(0) = \sum_i P_i \ket{\epsilon_i^0}\bra{\epsilon_i^0}$ is the projection of the initial density matrix onto the eigenstates of $H_S(0)$. 

Below we describe  two paradigmatic schemes that allow to experimentally obtain the work probability distribution with ancillas. 
The first one shows how to measure the characteristic function and the second one shows how to asses directly the work probability distribution while solely performing a measurement at the end of the transformation. Interestingly, both methods have been recently verified experimentally.

\subsection{Measuring the Characteristic Function: The Ramsey scheme}
\label{sec:ramsey}
In this section we review the first scheme, a Ramsey-inspired technique, to measure the work statistics. 
The method was originally proposed in Refs.~\cite{dorner2013extracting,mazzola2013measuring}, and gives access to the characteristic function by coupling the system to a two-level system (qubit).
The general idea of the scheme could be understood by noticing that the characteristic function for each $\lambda$ is given by the mean value of a unitary operator, as it is shown in Eq.~\eqref{eq:Gramsey}, 
and this mean value can be measured via an interferometric scheme.

The scheme starts with the system and the qubit in the product state $\rho_S(0)\otimes \ket{0}\bra{0}$ and we assume that $\rho_S(0)$ is diagonal in the initial energy basis, e.g. it is in a thermal state (we will discuss this assumption later). The Ramsey scheme consists of three steps:
\begin{enumerate}
\item The Hadamard gate \cite{nielsen-chuang_book} $H=(\sigma_z+\sigma_x)/\sqrt{2}$ is applied to the qubit. 
\item The qubit and the system  evolve according to the coupled evolution: 
\begin{eqnarray}
M_\lambda &=& U_S(\Time)e^{-i\lambda H_S(0)} \otimes \ket{0}\bra{0} +
\nonumber
\\
&+&e^{-i\lambda H_S(\Time)} U_S(\Time) \otimes \ket{1}\bra{1}.
\label{eq:mlambda}
\end{eqnarray}
\item The Hadamard gate $H$ is applied again to the qubit.
\end{enumerate}

At the end of the protocol the reduced density operator of the qubit is
\begin{equation}
\label{eq:rdmqubit}
\rho_D(\lambda) = \frac 12 \left [ I_Q + ({\rm Re} \chi_\lambda) \sigma_z 
+ ({\rm Im} \chi_\lambda) \sigma_y \right],
\end{equation}
where $I_Q$ is the identity on the auxiliary qubit.
Therefore, by repeating the experiments many times for a fixed value of $\lambda$, corresponding to the time of interaction between system and auxiliary qubit \cite{dorner2013extracting,mazzola2013measuring}, one can reconstruct the value of $\chi_\lambda$ through quantum state tomography of the qubit. Given the simple form of the density matrix \eqref{eq:rdmqubit},  only two measurements are needed: the mean values $\langle \sigma_z \rangle$, related to the qubit population imbalance, and $\langle \sigma_y \rangle$, related to the qubit coherence. To reconstruct the whole probability distribution $\Prob(W)$ one has to repeat the experiment many times for different values of $\lambda$ and then Fourier transform the characteristic function $\chi_\lambda$.

\begin{figure}[t]
\begin{center}
\includegraphics[width=\columnwidth]{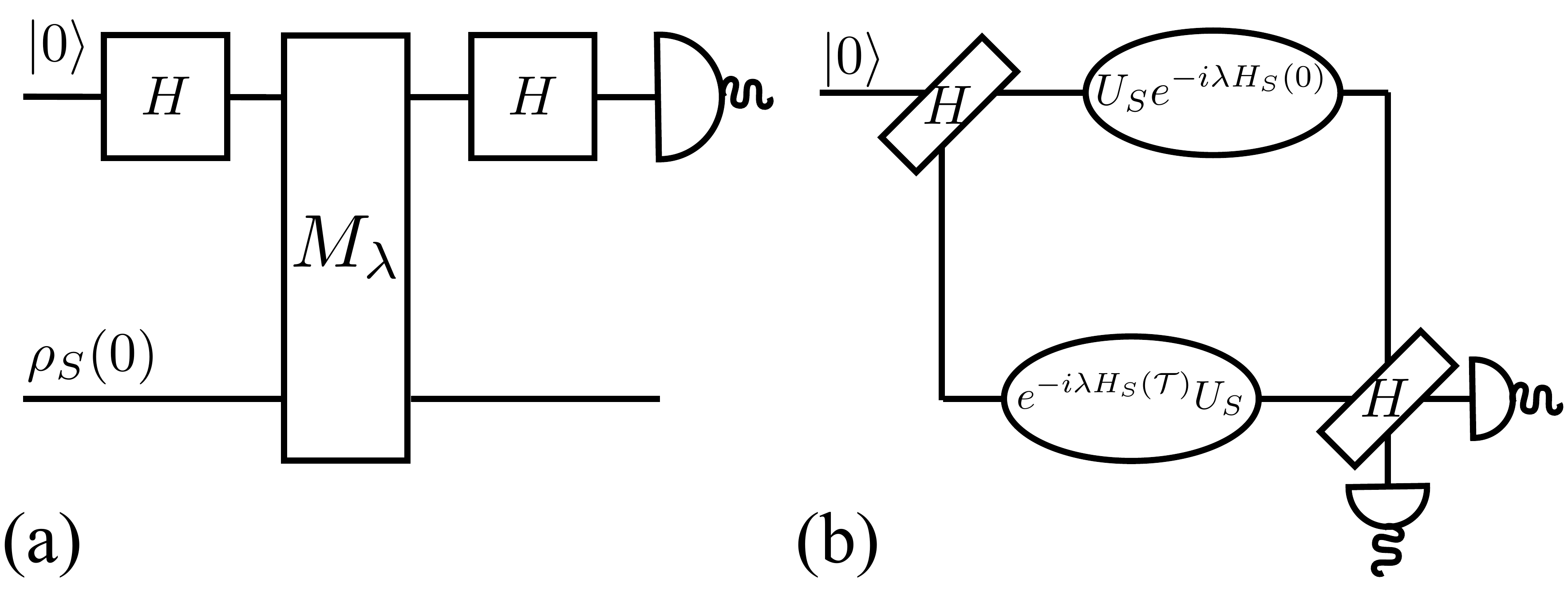}
\caption{(a) Quantum circuit realising the Ramsey scheme. The system is prepared in an initial state $\rho_S(0)$ which we assume diagonal in the initial energy basis. The qubit is initially in the state $\ket 0$. At the end of the evolution  the reduced density matrix of the qubit is reconstructed through state tomography.
 (b) Mach-Zehnder interpretation of the scheme. In the interferometer the state of the qubit is initially ``split" by a beam-splitter (corresponding to the Hadamard gate) in a superposition of $\ket 0$ and $\ket 1$ (corresponding to the two arms of the interferometer). In the upper arm of the interferometer the system is subject to the evolution $U_S(\Time)e^{-i\lambda H_S(0)}$ 
 while in the lower arm  $e^{-i\lambda H_S(\Time)} U_S(\Time)$. In the diagram for brevity we have written $U_S=U_S(\Time)$.
 At the end, the two arms pass through another beam-splitter (the second Hadamard gate) and the state of the qubit is reconstructed.
 }
\label{fig:scheme}
\end{center}
\end{figure}

The quantum circuit realising such a protocol is shown in Fig.~\ref{fig:scheme}(a) while a Mach-Zehnder interpretation of the scheme is presented in Fig.~\ref{fig:scheme}(b).

The advantage of the Ramsey scheme is that it bypasses the measurement of the initial and final energy of the system. The price to pay, however, is the need of implementing the conditional evolution operator $M_\lambda$.
We will discuss the implementation of the Ramsey scheme in Sec.~\ref{sec:physical}.

\subsection{Work Measurement with a single POVM}
\label{sec:POVM}

In this section we review the ideas developed in Ref.~\cite{roncaglia2014work} where it is shown that work, defined by the \tmp, 
can be measured using a single projective measurement over an ancillary system. Thus, work can be defined by a POVM, the most general possible quantum measurement that can be defined \cite{peres2006quantum} and very common
in quantum metrology and quantum information.
A POVM can be expressed in terms of a set of operators $A_k$, where $k$ labels all
possible different outcomes of the measurement, such that $A_k \geq 0$, $\sum_k A_k = I$, and the
probability of obtaining outcome $k$ of a quantum state $\rho$ is given by
$P(k) = \Trace(\rho A_k)$. Notice that the number of outcomes could in principle be larger than the dimension of the system.

A physical interpretation of such a generalised measurement is given by
Neumark's theorem \cite{peres2006quantum} which establishes that any POVM can be realised as
a projective measurement on an enlarged system that evolves unitarily. In terms
of the work measurement, this means that the work probability distribution can
be obtained by coupling the system to an ancillary system (or measuring
apparatus), letting them to evolve unitarily and then performing a projective
measurement at the end of the evolution.  Indeed, there is no need to perform two
 projective measurements at different times over the system, as the operational definition of work suggests, 
instead a single measurement at the very end of the protocol over an ancillary system can reveal the value of work.

It is worth noting that such an implementation is not
unique, in fact there are infinitely many combinations of ancillas, unitary
evolutions and projective measurements that are capable of realising the same
POVM. In Ref.~\cite{roncaglia2014work} two such implementations were proposed which we now
briefly discuss.

\begin{figure}[t]
\begin{center}
\includegraphics[width=\columnwidth]{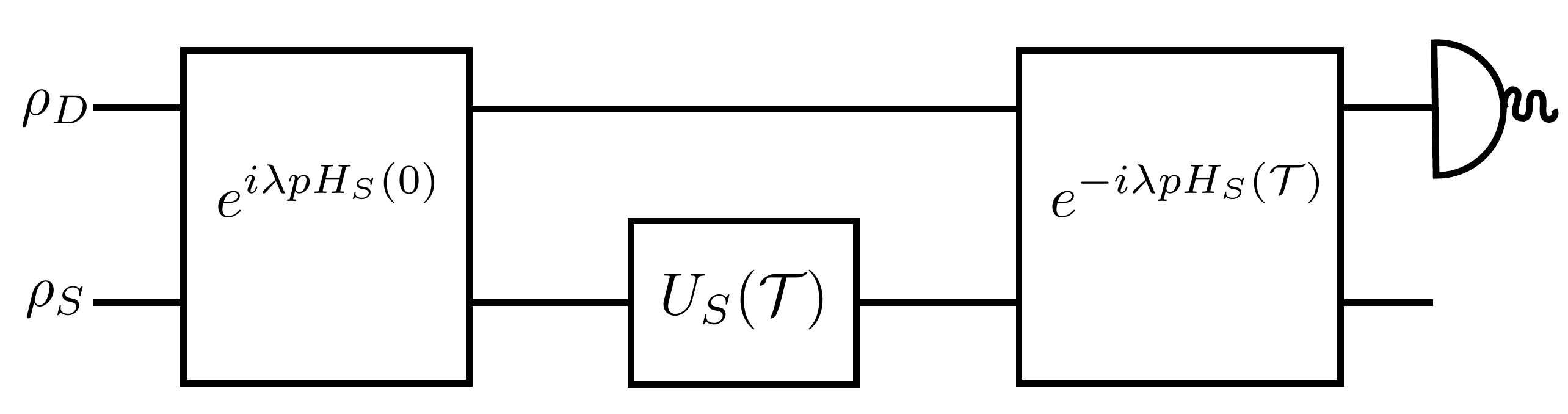}
\caption{Quantum circuit that measures work using a single POVM. This is a possible realisation where the detector ($D$) is 
a continuous variable system.  The two entangling operations apply a conditional
translation to the state of the detector. These translations at two different times depend on the value of the energy, thus the detector keeps
a coherent record of the value of work before the measurement.
 }
\label{fig:circPOVM}
\end{center}
\end{figure}

The POVM is implemented by introducing an auxiliary system or detector $D$ that interacts
with the system $S$ keeping a \emph{coherent record} of its energy at
two different times. The simplest strategy would be to consider $D$ as a continuous variable system (such that the operator
$p$ is the generator of translations) following these steps  \cite{roncaglia2014work}:

\begin{enumerate}
\item Initially $S$ and $D$ are in a product state, and $D$ could be, for instance, in  a position eigenstate or
a localized Gaussian state \cite{Cerisola2017,DeChiaraNJP2015}. 
\item $S$ and $D$ interact via an entangling evolution during time $\tau$ given by the Hamiltonian: 
$H_{SD}=-(\lambda/\tau)\,p\, H_S(0)$.
\item $S$ evolves according to a given process characterized by a unitary $U_S(\mathcal T)$.
\item $S$ and $D$ interact via an entangling evolution during time $\tau$ given by the Hamiltonian: 
$H_{SD}=(\lambda/\tau)\,p\, H_S(\mathcal T)$.
\item Projective measurement over $D$.
\end{enumerate}
In the protocol, $\lambda$ is some constant characterizing the strength of the interaction. 
In summary, the unitary sequence $U_\lambda=U_\lambda(\mathcal T)U_S(\mathcal T)U_\lambda^{\dagger}(0)$ is applied
before the measurement, with 
\begin{equation} \label{eq:uent}
U_\lambda(t)=e^{-i \lambda p H_S (t)}. 
\end{equation}
In Fig. \ref{fig:circPOVM}  it is shown the quantum circuit representing
the procedure. As an example, let us consider
an initial pure state of system and detector: $\ket{\psi_i}=\ket{0}_D\ket{\xi}_S$ after the application of $U_\lambda$ 
the systems will become correlated before the measurement in this way:
\begin{equation}
\ket{\psi_f} = \sum_{ij}\; \ket{\lambda\,\epsilon_{ji}}_D \;
 \left(\Pi_j^{\mathcal T} U_S(\mathcal T)\Pi_i^{0}\right)\ket{\xi}_S,
\end{equation}
where $\epsilon_{ji}=\epsilon_j^{\mathcal T}-\epsilon_i^{0}$  denotes a given value of work, $\ket{0}_D$  is an initial localised state, $\ket{\lambda\,\epsilon_{ji}}_D$ is the translated one, and $\Pi_j^{t}$ is the projector over the 
energy eigenspace of the Hamiltonian $H_S(t)$ with energy $\epsilon_j^{t}$. Notably, the detector keeps a coherent record of the 
value of work. Thus, all the information about the work distribution is encoded in this correlated final state.
In fact, a measurement that discriminates between the different states $\ket{\lambda\,\epsilon_{ji}}_D$ will reveal the value
of work $\epsilon_{ji}$ with a probability given by Eq. \eqref{eq:Pwtmp}. After many runs of the experiment it would
be possible to reconstruct the work probability distribution over this process.

The scheme allows different initial states for the system and detector, and also the measurement should be specified.
In fact, if one considers initial Gaussian states for the detector and a measurement in position, one could
set the interaction strength  $\lambda$ and the initial delocalisation of the state in order to 
obtain the work probability distribution \cite{DeChiaraNJP2015,talkner2016,Cerisola2017}.  Additionally, one can show that
initial coherences in the energy eigenbasis could affect the resulting work distribution, if the interaction strength $\lambda$ and the delocalisation
are not adjusted properly  (see Sup. Mat. \cite{Cerisola2017}).
On the other hand, the POVM can also be performed if one considers a discrete ancillary system. The
strategy is a variation of the phase estimation algorithm and was studied in \cite{roncaglia2014work}.
In this case, the measured distribution is a coarse grained
version of the real one that can be written as the convolution of $P(W)$ with a
windowing function whose precision depends on the dimension of the ancilla.

As mentioned before, the advantage of the POVM approach is that it does not
require the two time projective measurements and it furthermore allows one to
directly sample the work probability distribution. This is however done at the cost of having to implement appropriate
interactions between the system and the ancilla at different times. In Sec.~\ref{sec:physical} we review an experiment where this POVM was successfully implemented.

\section{General framework to measure quantum work distributions} \label{sec:general_framework}

In this section, we discuss a more general measurement framework to unify the two schemes discussed in the previous section.
The key point is to explicitly take into account the presence and the interaction between the system and the quantum detector used to store the information about work. 
This analysis thus provides a strong connection between energy fluctuations and the coherence of the initial system. As mentioned in the introduction we cannot always obtain a probability distribution of work when the initial state is endowed with quantum coherences~\cite{Perarnau-Llobet2017}. We will show that for the protocols described in this section, we either obtain a quasi probability distribution or a coarse-grained one.

\subsection{Dynamics of the system and the detector}

As discussed in the Introduction, the only clear way to define the work distribution of a quantum system is to give an operative implementation of the way we extract the information or, in other words, how we measure the energy invested or extracted in a quantum process. If we want to preserve  quantum effects yet be able to assess the energetic balance of a system, one way is to use a quantum detector or measurement apparatus. We thus consider the set-up in which a system $S$ is coupled to a detector $D$. This set-up is general enough to include most of the proposals in the literature and allows us to point out the differences and advantages of different measurement protocols.

As an illustrative detector we use a free quantum particle. The detector Hamiltonian is  $H_D = p^2/(2 m)$ where $p$ and $m$ are the momentum and the detector mass, respectively.
The coupling between $S$ and $D$ is obtained by turning on and off the interaction Hamiltonian $H_{SD}(t) =  -\beta(t) p  H_S(t)$.
This choice of the detector and the system-detector coupling are not limiting since, as discussed below, they account for all the relevant features and the schemes proposed in the literature.
At the same time, they allow us to simplify the discussion.
We will discuss also the changes needed in case other quantum detectors are used.

To determine the system internal energy variation, the time-dependent coupling strength $\beta(t)$ is chosen so that the detector ``records'' the energy of the system at the beginning and at the end of the evolution: $\beta(t)= \lambda/p_0 [ \delta(t-\Time) - \delta(t)]$, where the constant parameters $\lambda$ and $p_0$ have the dimension of time and momentum, respectively.

The delta-like behaviour of $\beta(t)$ must be considered as a limiting case in which coupling takes place over a time scale that is fast compared to the dynamics induced by $H_S(t)$ and $H_D$, so that the dynamics of the system and the detector are effectively ``frozen'' during the coupling. 
More specifically, if the interaction occurs in a time $\Delta t$, e.g., for $t \leq t' \leq t+\Delta t$, if $H_S(t)$ changes slowly in $\Delta t$, we can consider $H_S(t') \approx H_S(t)$.
Since $[H_S(t), H_{SD}(t)] = [H_D, H_{SD}(t)]=0$,  the full unitary operator can be easily calculated (setting $\hbar=1$)
\begin{eqnarray}
U(\Delta t)&=& e^{-i  \int_t^{t+\Delta t} dt' [H_S(t) - \lambda \beta(t) p  H_S(t) + \frac{p^2}{2 m}]  } \nonumber \\
&=& e^{-i  H_S(t) \Delta t} e^{i \lambda B p  H_S(t)}  e^{-i \frac{p^2}{2 m} \Delta t  } \nonumber \\
&\approx&  e^{i \lambda B p  H_S(t)}
\label{eq:system_detector_coupling}
\end{eqnarray}
where $B = \int_t^{t+\Delta t} dt' \beta(t)$ and the last step is obtained for small enough $\Delta t$.
We can assume that the coupling function $\beta(t)$ is normalised such that $B =1$. Then the dynamics during the system-detector coupling is given by $\exp{[i \lambda p  H_S(t)]}$, that is the one expected for a delta coupling.

The sequence of operations is the following: $i)$ at $t=0$, we turn on the system-detector coupling  that generates the dynamics described by $e^{-i    \frac{p \lambda }{p_0}  H_S(0)}$ as in Eq.~\eqref{eq:system_detector_coupling}, $ii)$ for $0\leq t \leq \Time$, we let the system and the detector evolve uncoupled with an external drive acting on the system. The corresponding unitary evolution is $U_S(\Time) U_D(\Time) = \Forder e^{-i \int_0^\Time dt H_S(t)} e^{-i  \frac{p^2}{2m} \Time}$ where we indicated with $ \Forder$ the temporal ordering. $iii)$ at $t=\Time$, we couple the system and the detector again to generate the dynamics $e^{i    \frac{p \lambda }{p_0}  H_S(\Time) }$ analogously to Eq.~ \eqref{eq:system_detector_coupling}.
The full evolution is described by the operator
$U_\lambda(\Time)= e^{i    \frac{p \lambda }{p_0}  H_S(\Time) }  U_S(\Time) U_D(\Time)   e^{-i    \frac{p \lambda }{p_0}  H_S(0)}$ \cite{solinas2015fulldistribution}.

Notice that neither the system nor the detector is projectively measured between times $0$ and $\mathcal T$.
The specific choice of $H_{SD}$ ensures that the coupling to the detector  does not induce any transition between the instantaneous system eigenstates.
In other words, system and detector do not exchange energy during the coupling.

We reasonably assume that system and detector are initially prepared in a separable and pure state (the assumption on initial purity can be relaxed for the system \cite{solinas2015fulldistribution}). 
Denoting with $\{ \ket{\epsilon_i^t}\}$ and $\{ \ket{p}\}$ the instantaneous eigenbasis  of $H_S(t)$ and the detector momentum basis, respectively, the more general initial state reads  $\ket{\phi_0}  = \sum_i  \int d p~ \psi_i^0 G(p) \ket{\epsilon_i^0, p}$.

By applying the evolution operator $U_\lambda(\Time)$  to $\ket{\phi_0}$ following the steps $i)-iii)$, we obtain 
\begin{widetext}
\begin{eqnarray}
  \ket{\phi_0 } &\rightarrow& \sum_{i} \int dp \psi_i^0 G(p) e^{- i \frac{p \lambda \epsilon_i^0}{p_0} } \ket{\epsilon_i^0, p} 
  \rightarrow \sum_{ij} \int dp \psi_i^0 G(p) e^{- i [\frac{p \lambda \epsilon_i^0}{p_0} + \frac{p^2}{2m} \Time ]}  U_{S,ji} \ket{\epsilon_j^\Time, p} \nonumber \\
  & \rightarrow & \sum_{ij} \int dp \psi_i^0 G(p) e^{i [\frac{p \lambda \epsilon_{ji}}{p_0} -\frac{p^2}{2m} \Time] } U_{S,ji} \ket{\epsilon_j^\Time, p} 
   \label{eq:phi_T}
\end{eqnarray}
\end{widetext}
where $U_{S,ji} = \matrixel{\epsilon_j^\Time}{U_S}{\epsilon_i^0}$ is the probability amplitude to go from $\ket{\epsilon_i^0}$ to $\ket{\epsilon_j^\Time}$ and, as before, $ \epsilon_{ji}= \epsilon_{j}^\Time -\epsilon_i^0$.

The corresponding final density matrix reads
\begin{eqnarray}
\rho_\Time = \sum_{ijkl} \int  dp dp' \rho_{ik}^0  G(p) G^*(p') U_{S,ji}  U_{S,kl}^\dagger  ~ && \nonumber \\
 \times  e^{i [\frac{p \lambda \epsilon_{ji}}{p_0} -\frac{p^2}{2m} \Time]} e^{-i [\frac{p'  \lambda \epsilon_{jk}}{p_0} -\frac{p'^2}{2m} \Time]}  \ketbra{\epsilon_j^\Time , p}{\epsilon_l^\Time , p'} &&
 \label{eq:rho_tot_p}
\end{eqnarray}
where $U_{S,kl}^\dagger= \matrixel{\epsilon_k^0}{U_S^\dagger}{\epsilon_l^\Time}$.
In writing the above equation, we have also implicitly defined the matrix elements of the initial density operator: $\rho_{ik}^0 =\matrixel{\epsilon_i^0}{\rho_S(0)}{\epsilon_k^0} =  \psi_i^0 (\psi_k^0)^*$.

The work done, i.e., the internal energy variation $ \epsilon_{ji} $, is now encoded in the detector degrees of freedom.
Therefore, we focus on the detector and trace out the system degrees of freedom.
The detector density operator $\rho_D(\Time) = \Trace_S [\rho_\Time ] = \sum_j \matrixel{\epsilon_j^\Time }{\rho_\Time }{\epsilon_j^\Time }$ reads
\begin{eqnarray}
\rho_D(\Time) &=& \sum_{ikj} \int  dp dp' \rho_{ik}^0 G(p) G^*(p') U_{S,ji} U_{S,kj}^\dagger  \nonumber \\
 &&e^{i [\frac{p \lambda \epsilon_{ji}}{p_0} -\frac{p^2}{2m} \Time]} e^{-i [\frac{p' \lambda \epsilon_{jk}}{p_0}-\frac{p'^2}{2m} \Time]} \ketbra{ p}{p'}.
\label{eq:rho_p_basis1}
\end{eqnarray}
As we can see, the contribution $\epsilon_{ji}$ appears in the phase accumulated between momentum eigenstates.
There is also an additional term  $\exp{ \{ -i p^2\Time/(2m) \}}$ due to the internal dynamics of the detector.
Since its value is known, we could subtract its contribution from the measured work distribution by data analysis.
However,  in many cases, it is convenient to cancel it during the protocol.
This step allows us also to keep the following discussion simple.

For a free particle detector we need to make a few additional assumptions.
We suppose that there are no degenerate states in the systems, i.e., $\epsilon_{ji} \neq 0$ for $i=j$, and that the initial detector state is a Gaussian function centered in $p=0$: $G(p) = [ \sigma^2/(2\pi)]^{1/4}~\exp{(-\sigma^2 p^2/4)}$ (the reason for the choice of the variance is discussed below). 
In this case, only states with momentum smaller than $p_{max} = 3/(2\sigma)$ are important in the dynamics since the others are exponentially suppressed.
Therefore, if we select $\lambda$, $m$, and $\Time$ in order to have $p_{max}^2\Time/(2m) = 9 \Time/(8 m \sigma^2)  \ll 1$, in Eq.~\eqref{eq:rho_tot_p} we can approximate $\exp{[i (\frac{p \lambda \epsilon_{ji}}{p_0} -\frac{p^2}{2m} \Time)]} \approx \exp{(i p \lambda \epsilon_{ji}/p_0)}$ (see also Refs.~\cite{talkner2016,Cerisola2017}).

Notice that, in general, the procedure to eliminate the dynamical contribution depends on the properties of both the detector and the initial state chosen.
For example, for a two-level quantum detector, the dynamical phase can be eliminated by additional operations as done in optics and nuclear magnetic resonance experiments \cite{jones2000, Ekert2000}. 
In this case, the dynamical phase accumulated by the two detector states $\ket{0}$ and $\ket{1}$ can be eliminated by swapping the states, i.e., $ \ket{0} \leftrightarrow \ket{1}$ and let the detector evolve for a time $\Time$.
At the end of this additional evolution the two states accumulate the same global phase that can be disregarded \cite{jones2000, Ekert2000}.
This phase elimination scheme can be applied also to the particle detector when it is initially in a superposition of two momentum eigenstates $\ket{p}$ and $\ket{p'}$. At the end of the evolution we can apply a unitary operator that switches the momentum states, i.e., $\ket{p} \leftrightarrow \ket{p'}$, let the detector evolve for time $\Time$ and neglect the overall dynamical phase.
In the following, we assume that the dynamical phase is cancelled or is negligible.

Under this approximation, the detector density matrix reads
\begin{eqnarray}
\rho_D(\Time) &=& \sum_{ikj} \int  dp dp' \rho_{ik}^0 G(p) G^*(p') U_{S,ji} U_{S,kj}^\dagger  \nonumber \\
 &&e^{i \frac{p \lambda \epsilon_{ji}}{p_0}} e^{-i \frac{p' \lambda \epsilon_{jk}}{p_0}} \ketbra{ p}{p'}.
\label{eq:rho_p_basis}
\end{eqnarray}
We can rewrite it in terms of the eigenstates $\ket{x}$ of the position operator $x$.
This can be done inserting the completeness operator $\int dx \ketbra{x}{x} = I$ and using the relation $\braket{x|p} = e^{i x p}/\sqrt{2 \pi}$.
We obtain 
\begin{eqnarray}
  \rho_D(\Time) &=& \sum_{ikj} \int  dx dx' \rho_{ik}^0 g\left(x+\frac{\lambda \epsilon_{ji} }{p_0}\right) \nonumber \\ 
 && \times g^*\left(x'+ \frac{\lambda\epsilon_{jk}}{p_0}\right)  U_{S,ji} U_{S,kj}^\dagger  \ketbra{x}{x'}
 \label{eq:rho_D_x}
\end{eqnarray}
where $g(x) = \frac{1}{\sqrt{2 \pi}} \int d p \exp{\{ i p x \} } G(p)$ is the Fourier transform of $G(p)$. 

With Eqs. \eqref{eq:rho_p_basis} and \eqref{eq:rho_D_x} in mind we can discuss different measurement schemes while maintaining the coupled system-detector evolution.
Since the detector momentum $p$ is a conserved quantity during the evolution,  the coupling between the system and the detector cannot induce transitions between different eigenstates of the momentum; instead, it changes their relative phase.
This picture is obviously reversed when considering eigenstates of the position, as the system-detector coupling $H_{SD}(t)$ induces transition between eigenstates of the position operator $x$.
This immediately suggests that there are two ways to extract the information about $\epsilon_{ji}$: we can either measure the momentum $\{ \ket{p} \}$  or the position $\{ \ket{x} \}$ of the detector.
Based on these considerations, we have the following two protocols to measure the work distribution.
In  {\bf Protocol $1$} we prepare the detector in a superposition of momentum eigenstates and measure their relative phase at the end of the evolution.
In  {\bf Protocol $2$} we prepare the detector in a position eigenstate and make a position measurement at the end of the evolution.
These protocols resemble the full-counting statistics formalism \cite{nazarov2003full, clerk2011full, hofer2016, Perarnau-Llobet2017} and the standard von Neumann measurement scheme \cite{VonNeumann1955}, respectively.
We now discuss in details their implications.

\subsection{Protocol $1$}
\label{sec:protocol1}

We first consider Protocol $1$ and focus on the phase accumulated between the momentum eigenstates $\ket{p_0/2}$ and $\ket{-p_0/2}$. 
The accumulated phase is given by $ \matrixel{p_0/2 }{\rho_D(t) }{-p_0/2 }$.
Rescaling for the initial phase $\matrixel{p_0/2 }{\rho^0_D }{-p_0/2 }$, we obtain the characteristic function of work~\cite{solinas2015fulldistribution} $\GFunc_\lambda=  \matrixel{p_0/2 }{\rho_D(t) }{-p_0/2 }/\matrixel{p_0/2 }{\rho^0_D }{-p_0/2 }$ where the $\lambda$ dependence is implicit in the dynamics of $\rho_D(t)$ [see Eqs. \eqref{eq:rho_p_basis} and \eqref{eq:rho_D_x}].

Notice that, with the choice of the initial states with $\pm p_0/2$, the detector dynamical phase does not contribute to $\matrixel{p_0/2 }{\rho^0_D }{-p_0/2 }$ as it can be seen directly from Eq.~ \eqref{eq:rho_p_basis}.
Since this results holds at all the orders, the constraint on the  variance of the initial detector states is not needed for the implementation of Protocol $1$.

The $\GFunc_\lambda$ function of the work reads \cite{solinas2015fulldistribution,solinas2016probing}
\begin{widetext}
\begin{eqnarray}
  \GFunc_\lambda = {\rm Tr}_S \Big[   U_{\lambda}(\Time) \rho_S(0) U^\dagger_{-\lambda}(\Time)  \Big] =
  {\rm Tr}_S \Big[  e^{i    \frac{p \lambda }{p_0}  H_S(\Time) }  U_S(\Time) e^{-i    \frac{p \lambda }{p_0}  H_S(0)} \rho_S(0) 
  e^{-i    \frac{p \lambda }{p_0}  H_S(0) }  U^\dagger_S(\Time) e^{i    \frac{p \lambda }{p_0}  H_S(\Time)} \Big].
  \label{eq:G_lambda}
\end{eqnarray}
\end{widetext}
and it is similar to the ones used in literature \cite{nazarov2003full, solinas2015fulldistribution, hofer2016, solinas2016probing, solinas2017measurement}. 
However, we must stress that we have not made any assumption about the initial state of the system.
If the system is initially in a thermal equilibrium state, i.e. $\rho_S(0) = \exp{\{ -\beta H_S(0) \}}/Z(0)$ where $Z(0)$ and $\beta$ are the partition function and the inverse temperature, respectively, $\exp{\{-i    \frac{p \lambda }{p_0}  H_S(0) \}}$ and $\rho_S(0)$ commute.
This allows us to simplify $\GFunc_\lambda$ and obtain the results of Ref. \cite{talkner2007fluctuation}.
Despite the similarities in the approaches and in the expressions, this characteristic function differs from the one in Eq. \eqref{eq:Gramsey}.
This is due to the fact that the two system-detector coupling schemes are different.
In any case the derivatives of $\GFunc_\lambda$ still gives information about energy fluctuations which do not necessarily coincide with the statistical moments of work.

A connection indeed exists for the first two moments. By direct calculation we obtain for the first two derivatives 
\begin{eqnarray}
 && (-i) \left . \frac{d \mathcal{G}_\lambda(p)}{d  p}  \right |_{\lambda=0} = 
\nonumber \\
&=&
 {\rm Tr}_S \Big[ H_S(\Time) \rho_S(\Time) - H_S(0) \rho_S(0) \Big].
 \label{eq:average_W}
\end{eqnarray}
and, using the property of the trace, 
\begin{eqnarray}
  &&(-i)^2 \frac{d^2 \mathcal{G}_\lambda(p)}{d \lambda^2}  \Big|_{\lambda=0}  =
  \\
  &=& {\rm Tr}_S \left\{ [ U^\dagger_S(\Time) H_S(\Time) U_S(\Time) - H_S(0)]^2 \rho_S(0) \right\}.
  \nonumber
\end{eqnarray}
These can be interpreted as the average work $\average{W}= \average{H_S(\Time)} - \average{H_S(0)}$ and its second moment \cite{engel2007jarzynski,campisi2011colloquium}.

Physically, $\GFunc_\lambda$ is associated to the accumulated phase in the detector and it can be measured by quantum tomography \cite{Paris-Rehacek_book}.
The characteristic function is therefore the primary measurement outcome in Protocol $1$. From an operational point of view, the dependence of $\GFunc_\lambda$ on $\lambda$ can be retrieved by varying either the time or the strength of the system-detector coupling.

There is an important difference between Eqs. \eqref{eq:G_lambda} and \eqref{eq:average_W} and the corresponding ones obtained with the \tmp{} and usually discussed in literature \cite{esposito2009nonequilibrium,campisi2011colloquium, dorner2013extracting,mazzola2013measuring, Batalho2014, roncaglia2014work}.
To highlight the differences,  let us explicitly write $\langle W\rangle$ as obtained from Eq.~\eqref{eq:average_W} (see also \cite{solinas2015fulldistribution}): 
 \begin{eqnarray}
\langle W\rangle
 &=& \sum_{i} P_i \sum_j P_{i\rightarrow j} (\epsilon_{j}^\Time - \epsilon_{i}^0) \nonumber \\
 &&+\sum_{j, i \neq k }  \epsilon_{j}^\Time ~ \rho_{ik}^0  U_{S,ji} U^\dagger_{S,kj} \ , 
 \label{eq:W_from_G}
 \end{eqnarray}
where $P_i = \rho_{ii}^0 $.

The first term in \eqref{eq:W_from_G} is the same as in \tmp{} \cite{engel2007jarzynski, esposito2009nonequilibrium, campisi2011colloquium} and can be straightforwardly interpreted in terms of classical conditional probabilities.
In absence of initial coherences or when the first projective measurement in the \tmp{} destroys them, the system can be treated as a classical ensemble with probability  $P_i $ to start from  the state $\ket{\epsilon_i^0}$ and end up in the state $\ket{\epsilon_j^\Time}$ with conditional probability $P_{i\rightarrow j}$.
In this framework, the energy of the initial state is always well defined as it is the trajectory associated to the work $\epsilon_j^\Time - \epsilon_i^0$.

However, this is not the full story since the remaining terms, which depend on the initial coherences $\rho_{ik}^0$, are of a purely quantum nature. These coherences are destroyed by the initial measurement of $H_S(0)$ performed in \tmp{}. 
This changes the dynamics as discussed in the introduction and, using Feynman's words \cite{feynman1965quantum}, this destroys all the interference alternative of the dynamics.

To fully understand the impact of the initial coherences on the quantum work, we must answer to a much more subtle question: can the Fourier transform of the observable function $\GFunc_\lambda$ be interpreted as a probability distribution of the work?
Classically, we can obtain the work probability distribution as the Fourier transform of the corresponding characteristic function. However for initial coherent states the function: $\Prob_{\rm quasi}(W)=1/(2\pi)  \int d \lambda~\exp{\{ - i  \lambda W \}} \GFunc_\lambda $ is a quasi-probability distribution, i.e. it is real but not positive definite.
We can write \eqref{eq:G_lambda} explicitly using the $\{ \ket{\epsilon_j^\Time} \}$ and then calculate the Fourier transform.
The details of the calculations can be found in Refs. \cite{solinas2015fulldistribution,solinas2016probing}; we obtain
\begin{equation}
 \Prob_{\rm quasi}(W) =  \sum_{ikj} \rho_{ik}^0  U_{S,ji} U_{S,kj}^\dagger \delta \Big[W - \left(\epsilon_j^\Time- \frac{\epsilon_i^0+\epsilon_k^0}{2}\right) \Big].
 \label{eq:P(W)}
\end{equation}
If the density operator has no initial coherences, i.e., $\rho_{ik}^0=0$ for $i\neq k$, we obtain the work probability distribution Eq.~\eqref{eq:Pwtmp} of the \tmp{}.
Following this reasoning, we refer to the terms with $i=k$ in Eq. \eqref{eq:P(W)} as ``classical contributions''.

In general, however, $ \Prob_{\rm quasi}(W)$ depends on the additional off-diagonal density matrix terms which are the fingerprints of coherent quantum processes.
To strengthen this view, we notice that the contributions $\rho_{ik}^0  U_{S,ji} U_{S,kj}^\dagger$ are always real \cite{solinas2016probing} but  they are not constrained to be positive.
Therefore, due to the presence of these terms, $\Prob_{\rm quasi}(W)$ is not  positive definite and can only be referred to as a quasi-probability distribution \cite{solinas2015fulldistribution, allahverdyan2014, hofer2016, Perarnau-Llobet2017,Hofer2017quasiprobability}. 

This can be surprising, but it is not contradictory.
We stress once again that the measured quantity in Protocol $1$ is the characteristic function $\GFunc_\lambda$, i.e., the accumulated phase.
Being an experimentally measurable quantity, it is free of ambiguous interpretations.
The work quasi-probability distribution $\Prob_{\rm quasi}(W)$ obtained from Fourier transforming the accumulated detector phase $\GFunc_\lambda$ is instead a {\it derived} quantity that we would normally {\it associate} to a probability distribution.
The above discussion points out that this last step is not allowed if the system has initial coherences.

This situation is not new in quantum mechanics, in fact, it turns out to be associated to the violation of the  Leggett-Garg inequality \cite{Leggett1985, Emary2014, bednorz2010quasiprobabilistic, bednorz2012nonclassical, clerk2011full, solinas2015fulldistribution, Perarnau-Llobet2017}.
In the Leggett-Garg model the negativity of the probability  distribution occurs when we try to interpret the results of a quantum measurements in purely classical terms.
In our case, the macrorealism hypothesis \cite{Leggett1985, Emary2014} is violated since the system is initially in a coherent superposition of energy eigenstates and, thus, it does not have a well-defined energy.
Similar results and interpretations have been discussed for the full counting statistics as arising from quantum interference effects \cite{bednorz2010quasiprobabilistic, bednorz2012nonclassical, clerk2011full, hofer2016}.
By reversing this argument, we can consider the negative regions of $\Prob_{\rm quasi}(W)$ as signatures of pure quantum effects.

The important question is whether Protocol $1$ has any advantage over the \tmp{}.
The positive or negative answer depends on whether we have and want to preserve the quantum effects and dynamics.
Recalling the example discussed in the Introduction, with Protocol $1$ we can measure the work moments and, thus, the energy needed to run the database search algorithm without affecting the algorithm performance.

It is worth noticing that the idea behind Protocol $1$ can be extended to measure the dissipated heat of an open quantum system \cite{solinas2015fulldistribution}.

\subsection{Protocol $2$}

The second protocol is based on the position measurement of the detector at the end of the dynamical evolution.
The system-detector interaction proportional to $ p H_S$ generates a shift in the position $x$ that is proportional to the energy of the system.
The coupling at the beginning and at the end gives us information about the work done on the systems during the two couplings.
The relevant equation is \eqref{eq:rho_D_x} where we have explicitly written the detector position and its shifts.

To clarify the physical picture, let us discuss first the idealised case in which the detector is strongly localised in $x_0$: $g(x)=\delta(x-x_0)$.
Formally, this limit is not correct because of the assumption we have made to neglect the dynamical phase of the free particle detector.
Despite this, it allows us to point out an important feature of the protocol.
In addition, this situation can be realistic for other kinds of detectors, where we can have both localised initial states and can remove the dynamical phase (see above discussion on qubit detector and bosonic mode as in Ref. \cite{solinas2016probing}).

In this case, the density matrix of the detector in Eq. \eqref{eq:rho_D_x} reads 
\begin{eqnarray}
  \rho_D(\Time) &=& \sum_{ikl} \int  dx dx' \rho_{ik}^0  \delta \left(x-x_0-\frac{\lambda \epsilon_{ji} }{p_0}\right) \nonumber \\ 
  && \times \delta \left(x'-x_0- \frac{\lambda \epsilon_{jk} }{p_0}\right)  U_{S,ji} U_{S,kj}^\dagger  \ketbra{x}{x'}.  \nonumber
   \label{eq:P_delta_funcitons}
\end{eqnarray}
The probability to measure $\bar{x}$ and, thus, to have a shift in the detector of $\Delta x= \bar{x}-x_0$ is $ \Prob(\Delta x) =\matrixel{\bar{x}}{\rho_D}{\bar{x}}$.
Using Eq. \eqref{eq:P_delta_funcitons} and performing the integration, we find that the probability distribution is non-zero only if $\epsilon_{i}^0 = \epsilon_{k}^0$.
For non-degenerate cases, this imposes the additional constraint $k=i$ and the probability distribution reads
\begin{equation}
  \Prob(\Delta x) = \sum_{ij} P_i P_{i\rightarrow j} ~\delta \left[  \Delta x- \frac{\lambda }{p_0}(  \epsilon_{j}^\Time - \epsilon_{i}^0) \right].   \nonumber
  \label{app_eq:rho_T_gx}
\end{equation}
The latter result coincides with the classical contribution in Eq. \eqref{eq:P(W)} and the one obtained with the the \tmp{} \cite{kurchan2000quantum,tasaki2000jarzynski, esposito2009nonequilibrium, campisi2011colloquium}.
For a localised initial detector state, we obtain the classical work distribution.

A more realistic situation is the one in which the detector is not perfectly localised \cite{VenkateshNJP2014,WatanabePRE2014a,WatanabePRE2014b,sokolovski2015meaning, talkner2016}.
This uncertainty in the initial position affects the measurement of the work distribution.
Following the above discussion we consider an initial state in the momentum basis as $G(p) = [ \sigma^2/(2\pi)]^{1/4}~\exp{(-\sigma^2 p^2/4 + i p x_0)}$ so that for $\Time/(m \sigma^2) \gg 1$ and we can neglect the detector dynamical phase.
The corresponding initial state in the coordinate representation is $g(x)= \exp\{-\frac{(x-x_0 )^2}{4 \sigma ^2}  \}/ \sqrt[4]{2 \pi \sigma^2} $.
The probability to measure $\Delta x$ gives us the wok distribution as \cite{solinas2016probing}
\begin{eqnarray}
 \Prob(\Delta x) 
  &=& \sum_{ikj} \rho_{ik}^0 U_{S,ji} U_{S,kj}^\dagger 
  \nonumber
  \\
  &\times&
  \frac{e^{-\frac{\left(\Delta x- \frac{\lambda \epsilon_{ji} }{p_0}\right)^2 +\left(\Delta x- \frac{\lambda \epsilon_{jk} }{p_0}\right)^2 }{4 \sigma ^2}}}
  {\sqrt{2 \pi } \sigma }. 
  \label{eq:PDeltax}
  \end{eqnarray}
If the system has no initial coherences, i.e., $\rho_{ik}^0=0$ for $i\neq k$, the work distribution can be interpreted as a classical one with uncertainty \cite{solinas2016probing}.
An analogous situation is obtained when in presence of initial coherences the Gaussian functions in Eq. \eqref{eq:PDeltax} do not overlap.
This situation occurs when $\epsilon_j^\Time - \epsilon_i^0 \ll  \sigma p_0/\lambda$ (for any $j$ and $i$) and the off-diagonal contributions do not contribute to Eq. \eqref{eq:PDeltax}.
This corresponds to the physical case in which  the uncertainty in the initial state is small with respect to energy variation that we want to measure.
This allows us to perform a ``precise" measurement of the work and to distinguish the classical trajectories and evolution associated to the energy exchanges.

At the opposite limit we have an ``imprecise" measurement when $\epsilon_k^0 - \epsilon_i^0 \approx  \sigma p_0/\lambda$.
In this case, the Gaussian functions in Eq. \eqref{eq:PDeltax} overlaps and the off-diagonal contributions $\rho_{ik}^0 $ become important again.
Notice that in this case $\Prob(\Delta x)$ is still a true probability distribution, i.e., it is definite positive \cite{solinas2016probing}, since it is a direct measurement output.
In presence of a considerable uncertainty,
the classical and quantum distribution (with and without initial coherence $\rho_{ik}^0$) deviate one from each other.
Therefore, the uncertainty reveals the presence of an underlying coherent quantum dynamics that manifests itself as a modification of the work distribution \cite{solinas2016probing, talkner2016}.

\section{Physical implementations of the measurement schemes}
\label{sec:physical}

\subsection{Implementations of the Ramsey scheme}
In Sec.~\ref{sec:ramsey} we discussed the use of an auxiliary qubit coupled to the system to measure its work characteristic function during a dynamical process. In Sec.~\ref{sec:protocol1} the Protocol 1 was introduced as a generalisation of the Ramsey scheme.
Although the scheme eliminates the need for the initial and final measurement of the system energy, which in some physical systems can be challenging, the requirement is to couple the qubit and the system through the operator $M_\lambda$ (see Eq.~\eqref{eq:mlambda}).
In Ref.~\cite{mazzola2013measuring} it has been shown that the gate $M_\lambda$ can be decomposed as a series of local unitary transformations and proper controlled operations, i.e. gates that apply an operator to the system only if the qubit is in the state $\ket 1$. Indeed we have:
\begin{eqnarray*}
M_\lambda &=& (I_S \otimes \sigma_x) M_2(\lambda)(I_S \otimes \sigma_x)M_1(\lambda)
\\
M_1(\lambda)&=& I_S\otimes \ket 0\bra 0+e^{-i\lambda H_S(\Time)} U_S(\Time)
\otimes \ket 1\bra 1
\\
M_2(\lambda)&=& I_S\otimes \ket 0\bra 0+U_S(\Time)e^{-i\lambda H_S(0)} 
\otimes \ket 1\bra 1 .
\end{eqnarray*}
Moreover, Ref.~\cite{dorner2013extracting} employs a  protocol equivalent to the Ramsey scheme and shows that it only requires a conditional qubit-system interaction of the form 
$V\otimes \ket 1\bra 1 $ where $V$ is the operator that is varied in time during the work protocol.

Ref.~\cite{dorner2013extracting} describes an ion trap experiment to measure the work done in displacing spatially the ion trap. The displacement can be described by the operator $V=g(t) x$, where $x$ is the ion position. The auxiliary qubit is represented by two internal electronic levels of the ion. The required interaction between the qubit and the ion position is provided by illuminating the ion with a far-detuned elliptically polarised standing-wave laser field. In the Lamb-Dicke regime the interaction induced is of the form:
\begin{equation}
H_I(t) = x\otimes\left[g_0(t) \ket{0}\bra{0} +g_1(t)  \ket{1}\bra{1}\right ]
\end{equation}
where the couplings $g_i(t)$ can be easily controlled by the relative intensities of two orthogonal laser polarisations.  

Ref.~\cite{mazzola2013measuring} instead proposes two possible implementations of the Ramsey scheme with mechanical oscillators. In the first version, an optical cavity is realised with two mirrors, one of them vibrating and playing the role of the system. The cavity also contains a two-level atom realising the auxiliary qubit which modifies the cavity field which ultimately drives the vibrating mirror. The work done on or extracted from the latter is thus conditional on the state of the qubit which acts as a switch. Conversely, one can measure the state of the qubit to estimate the work done on the vibrating mirror. In the second scenario, the system is embodied by a suspended nanomechanical beam which is coupled capacitively to a Cooper-pair box realising the auxiliary qubit. Again the position of the vibrating beam is coupled to an operator of the qubit and allows one to measure the work done on the beam by an external driving. 

After the original proposals, other similar variations of the protocol have been put forward. In particular Ref.~\cite{campisi2013employing} extends the scheme to a circuit quantum electrodynamics setup for measuring the work done in a open driven system including a strongly dissipative regime. In Ref.~\cite{GooldPRE2014}, the authors propose a Ramsey scheme to assess the heat distribution in a quantum process and the connection to the Landauer principle. The authors of Ref.~\cite{JohnsonPRA2016} employ the Ramsey scheme to measure work done on a gas of ultracold atoms. This ultimately allows them to connect work fluctuations with fluctuation theorems, e.g. the Jarzynski equality, and estimate the temperature of the gas with a precision below one nanokelvin. 

\begin{figure}[t]
\begin{center}
\includegraphics[width=\columnwidth]{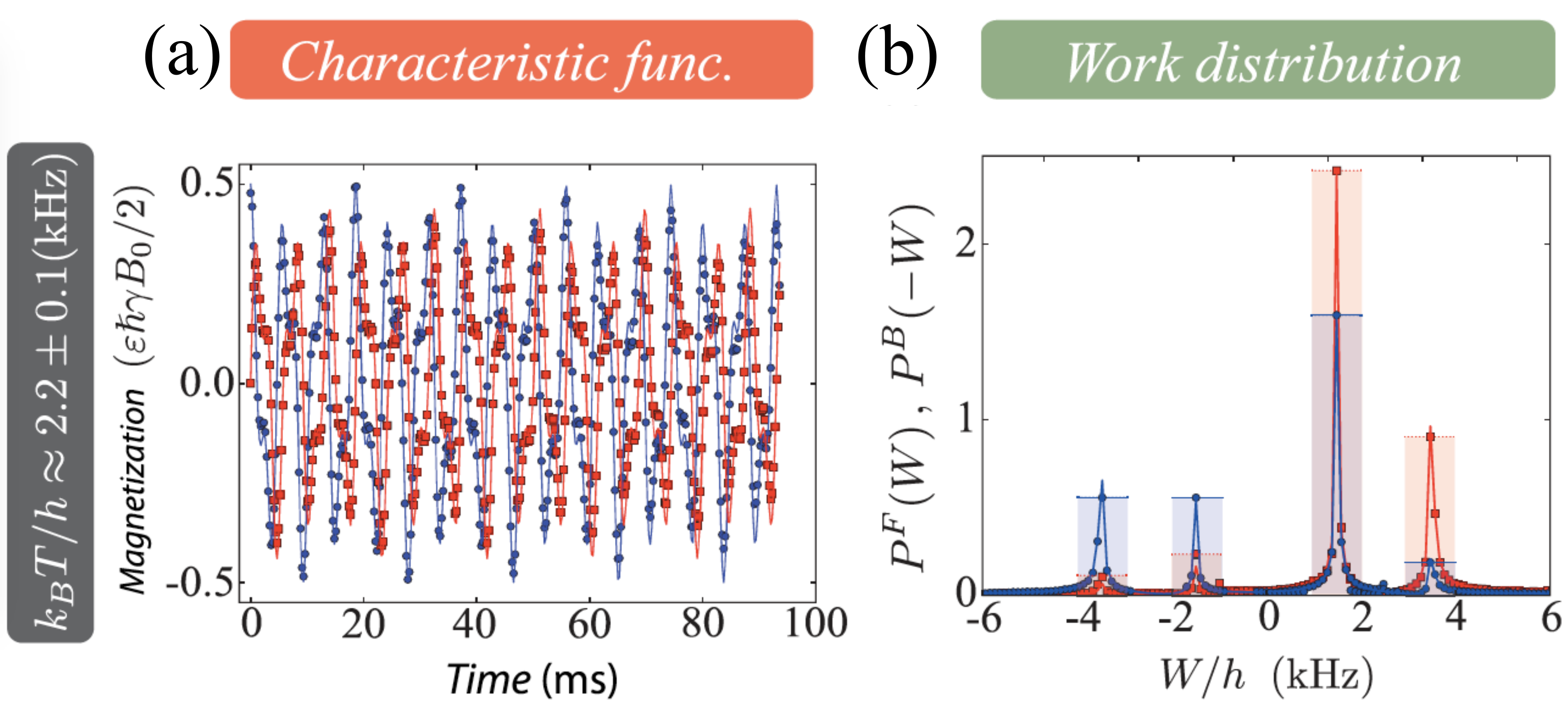}
\caption{(a) Experimental data for the magnetisation $\sigma_x$ and $\sigma_y$ of the ${}^1$H spin (blue circles and red squares, respectively), plotted against the time length $\lambda\pi\nu_2/J$. The solid lines show Fourier fittings, which are in agreement with the theoretical simulation of the protocol. (b) The experimental points for the work probability distribution corresponding to the forward (backward) protocol are shown as red squares (blue circles) and obtained from the Fourier transform of the data in panel (a). Figure adapted with permission from Ref.~\cite{Batalho2014}.
 }
\label{fig:nmr1}
\end{center}
\end{figure}

The Ramsey scheme was experimentally realised in 2014 in the nuclear magnetic resonance (NMR) experiment reported in Ref.~\cite{Batalho2014}, only a year after the original proposals. The experiment employs a liquid sample of chloroform molecules and involves an NMR-spectroscopy of the ${}^1$H and ${}^{13}$C nuclear spins of the molecule. The ${}^{13}$C spin plays the role of the driven system while the ${}^1$H spin acts as the auxiliary qubit. The goal is to measure the work performed on the carbon spin subject to a rotating magnetic field described by the Hamiltonian:
\begin{equation}
H(t)=2\pi\hbar\nu(t)\left(\sigma_x^C\sin\frac{\pi t}{2\tau}+\sigma_y^C\cos\frac{\pi t}{2\tau} \right)
\end{equation}
When the magnetic field intensity $\nu(t)$ is changed in a time $\tau$ with a linear ramp: $\nu(t)=\nu_1 +(\nu_2-\nu_1)t/\tau$. 
  In the above Hamiltonian $\sigma_i^C$ is the $i$-th Pauli operator acting on the C nuclear spin while the ones acting on the H nuclear spin are denoted with $\sigma_i^H$.
The process is performed both in the forward direction, increasing $\nu(t)$ from $\nu_1$ to $\nu_2>\nu_1$, and in the backward direction, by exchanging $\nu_1$ and $\nu_2$. The Ramsey protocol is implemented by taking advantage of the natural interaction between the carbon and hydrogen spins: $H_I = 2\pi J\sigma_z^H\sigma_z^C$ where $J$ is their coupling rate. The controlled gates $M_1$ and $M_2$ are realised using the free evolution induced by $H_I$ and correcting with single-qubit gates.

At the end of the process the magnetisation of the hydrogen spin is measured to reveal the work characteristic function. The state of the carbon qubit (the system) was initialised in a thermal state with different temperatures. The results for the work characteristic function and the work probability distribution obtained by a Fourier transform are presented in Fig.~\ref{fig:nmr1} with diagrams from Ref.~\cite{Batalho2014}. The real and imaginary parts of the characteristic function $\chi_\lambda$ exhibit well defined oscillations as a function of the coupling time $\lambda$ (measured in ms in the diagram) with a weak decay due to thermal relaxation which could potentially alter the work distribution. However the sampling time of the characteristic function has been kept much smaller than the relaxation time so that the effects of relaxation and dephasing are minimal. 

The Fourier transform of the characteristic function reveals a work probability distribution, shown in Fig.~\ref{fig:nmr1}(b) with four well defined peaks corresponding to the four allowed transitions between the two initial carbon spin states and the two final ones.

\begin{figure}[t]
\begin{center}
\includegraphics[width=\columnwidth]{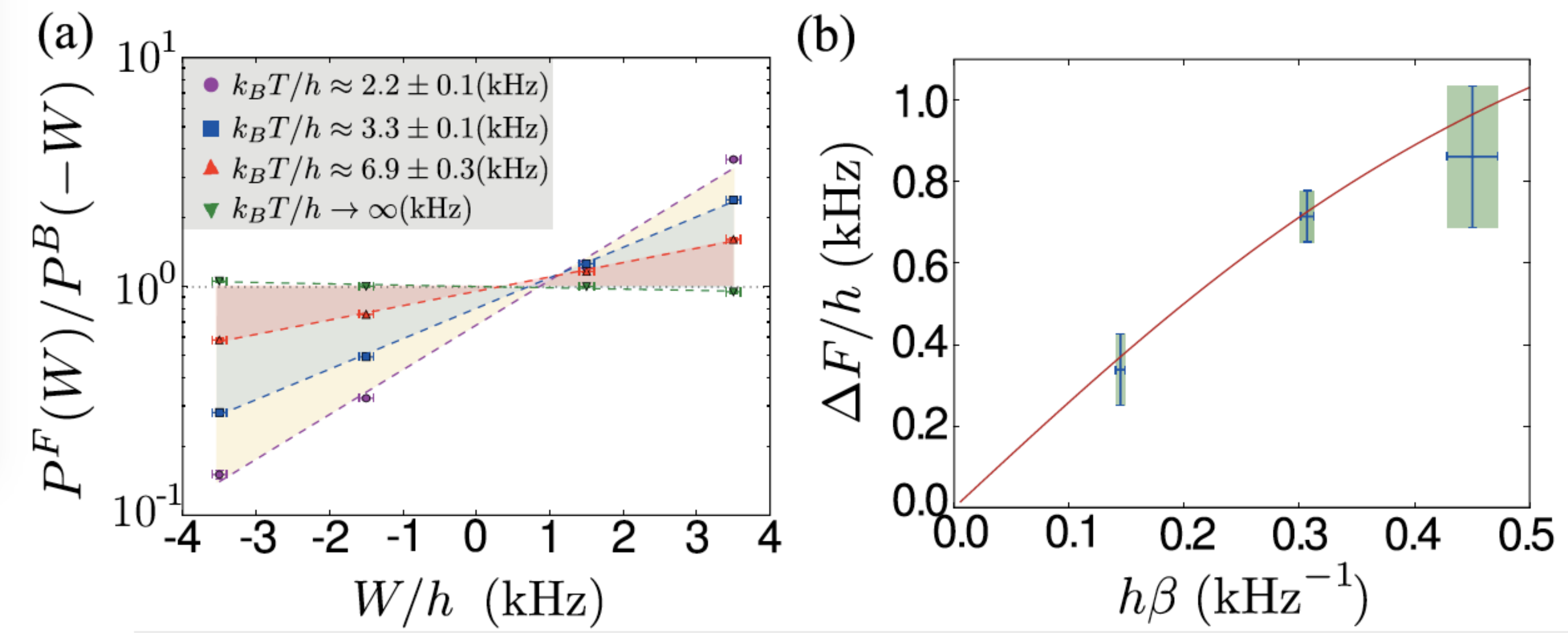}
\caption{(a) The ratio $\Prob^F(W)/\Prob^B(-W)$ is plotted in logarithmic scale for four values of the carbon spin initial temperature. The data are obtained from the values in Fig.~\ref{fig:nmr1}. (b) Mean values for the free energy difference $\Delta F$ and the initial inverse temperature $\beta$ obtained using a linear fit of the data corresponding to panel (a). The full red line represents the theoretical expectation, Eq.~\eqref{eq:theoreticaldeltaF}.
Figure adapted with permission from Ref.~\cite{Batalho2014}.
 }
\label{fig:nmr2}
\end{center}
\end{figure}

From the reconstructed $\Prob(W)$, the authors of Ref.~\cite{Batalho2014} verified for the first time the Tasaki-Crooks relation in a fully quantum regime. To this end in Fig.~\ref{fig:nmr2}, the ratio of the forward to the backward work probability distributions $\Prob^F(W)/\Prob^B(-W)$ is shown for four different initial temperatures. The logarithm of this ratio is a linear function of the work $W$ thus verifying the Tasaki-Crooks relation:
\begin{equation}
\ln \frac{\Prob^F(W)}{\Prob^B(-W)}= \beta(W-\Delta F) 
\end{equation}
where $\beta$ is the initial inverse temperature of the carbon spin and $\Delta F$ is the free energy difference between the final and the initial thermal equilibrium states. By fitting the numerical data for
$\Prob^F(W)/\Prob^B(-W)$, the authors of Ref.~\cite{Batalho2014}, verified that the free energy difference obtained from the Tasaki-Crooks relation coincides, within experimental errors, with the theoretical value:
\begin{equation}
\label{eq:theoreticaldeltaF}
\Delta F= \frac{1}{\beta} \ln\frac{\cosh(\beta\nu_1)}{\cosh(\beta\nu_2)}
\end{equation}
They also verified the Jarzynski equality by experimentally computing $\ln\langle e^{-\beta W}\rangle$ and comparing it to the free energy difference obtained from the fitting of the Tasaki-Crooks relation or from the theoretical value \eqref{eq:theoreticaldeltaF} finding excellent agreement for different initial temperatures. 

Before ending this section, it is worth commenting that the Ramsey scheme does not impose a restriction on the initial state of the system being thermal or diagonal in the initial energy eigenbasis. The Ramsey scheme allows one to measure the following quantity:
\begin{equation}
 {\rm Tr}_S \Big[   U_S^\dagger(\Time) e^{i\lambda H_S(\Time)}U_S(\Time) e^{-i\lambda H_S(0)}\rho_S(0)   \Big].
\end{equation}
where the initial state $\rho_S(0)$ can have coherences in the initial energy basis. In this case however this measured quantity does not correspond to a characteristic function of a probability distribution since its Fourier transform is in general complex. This is once more an indication of the presence of coherences in the initial state. 

\subsection{Measuring work using light as an ancilla}

In Sec. \ref{sec:POVM} we review the method proposed in \cite{roncaglia2014work} that directly measures work with a single POVM. 
This strategy removes the need to make two measurements of the energy at different times, at the cost of implementing two entangling operations.
An implementation with cold atomic ensembles was proposed in Ref.~\cite{DeChiaraNJP2015} which we here discuss briefly. 

The scheme allows one to reconstruct the work done on an atomic ensemble subject to a rotating magnetic field of constant amplitude $B$ and described by the Hamiltonian:
\begin{equation}
H(t) = -\gamma B \left[\cos\alpha(t) J_z + \sin\alpha(t) J_y\right]
\end{equation}
where $\gamma$ is the gyromagnetic ratio and we introduced the collective atomic angular momentum ${\bf J}=(J_x,J_y,J_z)$. To measure work, the authors of Ref.~\cite{DeChiaraNJP2015} propose the use of a continuous variable light mode embodied by the fluctuations of the polarisation along two orthogonal directions of a travelling laser mode and described by two conjugate position $X$ and momentum $P$ quadratures. The conditional shift needed to imprint the energy of the atomic system onto the light mode is provided by the Faraday rotation which couples the atomic angular momentum $\bf J$ with the light polarisation through the coupled transformation $U_I =\exp[-i\kappa P (\cos\phi J_z+\sin\phi J_y)]$ where the coupling constant $\kappa$ depends on the intensity of the beam light, on the atom-photon interaction and the angle $\phi$ depends on the direction of propagation of the light beam with the quantisation axis of the atomic ensemble. Notice that $U_I$ is spatial translation operator of the light mode analogous to the transformation generated by the system-detector Hamiltonian $H_{SD}$ of Eq.~\eqref{eq:uent}.

\begin{figure}[t]
\begin{center}
\includegraphics[width=\columnwidth]{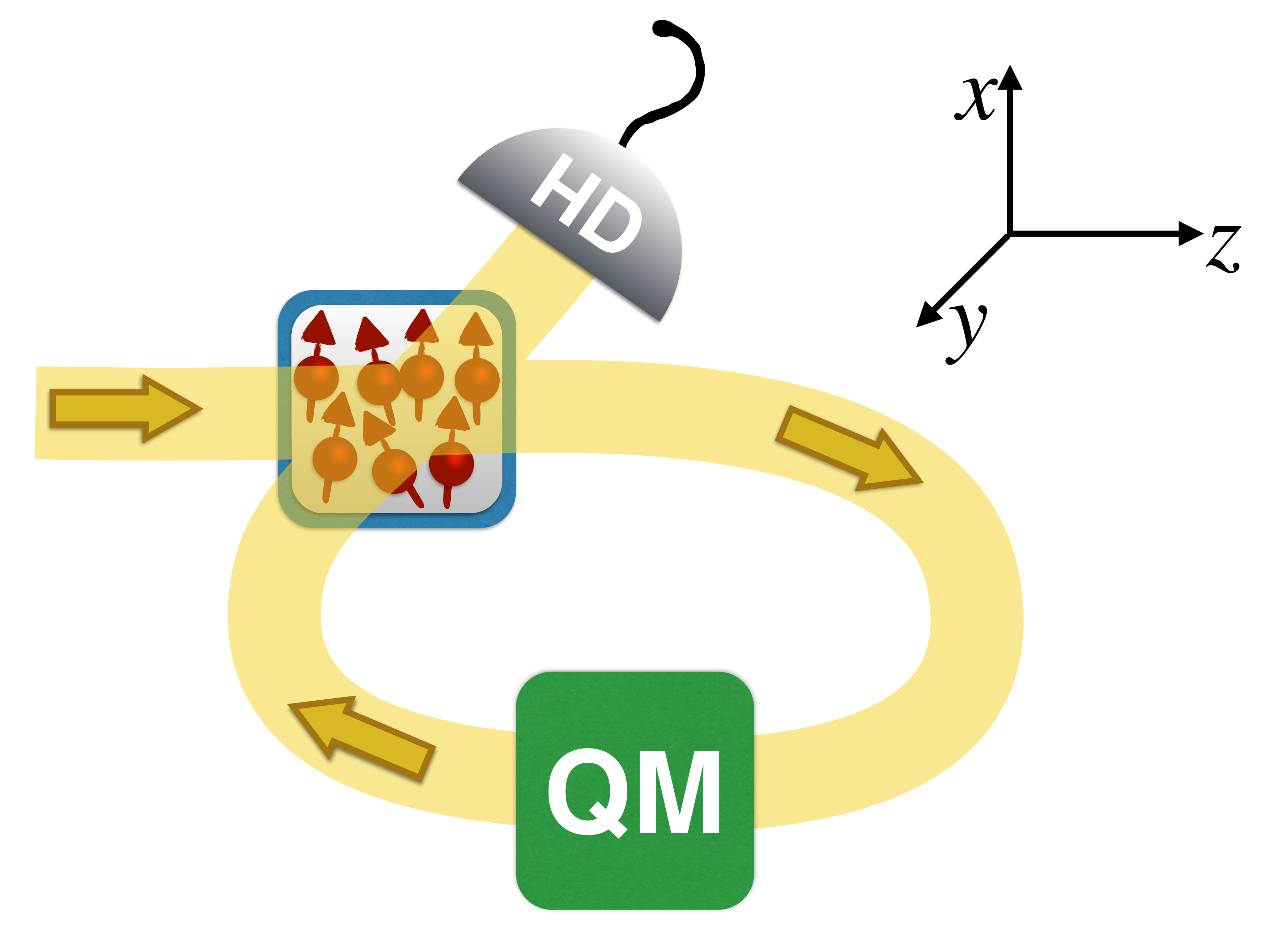}
\includegraphics[width=\columnwidth]{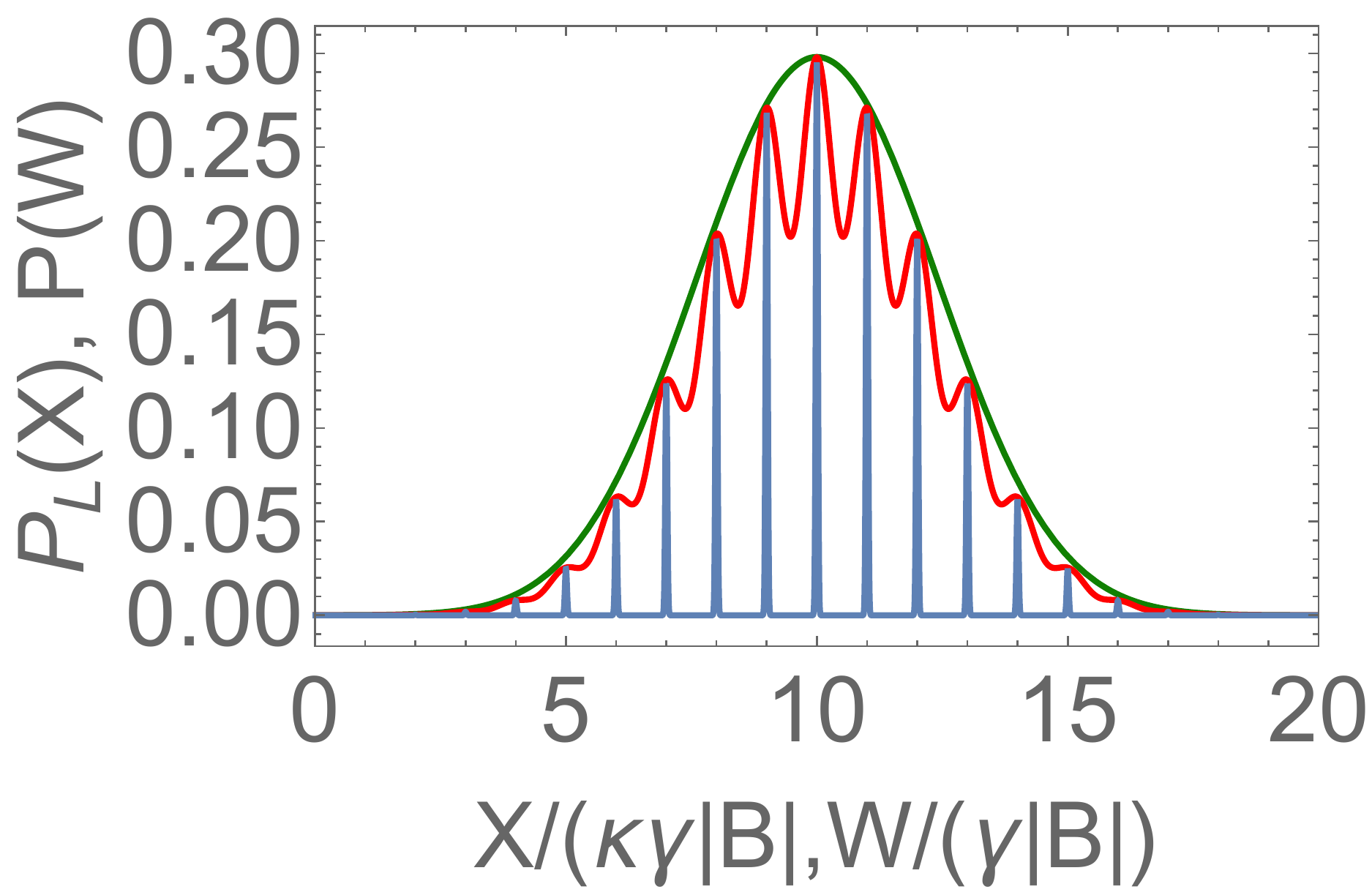}
\caption{Top: Setup to measure the probability distribution of the work done on an atomic ensemble. A polarised beam is sent along the $z\;\; (\phi=0)$ direction to interact with the atomic ensemble. The beam is then stored in a quantum memory (QM) while the magnetic field applied to the atoms is rotated from $z$ to $y$. Finally the beam is redirected to the atoms at an angle $\phi=-\pi/2$ and eventually measured. Figure reproduced with permission from Ref.~\cite{DeChiaraNJP2015}. 
Bottom: for an instantaneous rotation of the magnetic field, comparison of the work probability distribution (blue) with the rescaled light polarisation distributions for $N=20$ atoms, $\sigma^2=1/2$, $\kappa =2$ (red) and $\kappa=1$ (green).}
\label{fig:schemelight}
\end{center}
\end{figure}

The protocol to extract the work distribution consists in {\it i.} shining the laser at an angle $\phi=\pi+\alpha(0)$; {\it ii.}  while the light beam is stored in a quantum memory, performing the rotation of the magnetic field from an angle $\alpha(0)$ to an angle $\alpha(\Time)$; {\it iii.} finally, letting the light beam to pass again through the atomic sample at an angle $\phi=\alpha(\Time)$. At the end of the process, the light beam is analysed through standard homodyne detection to reconstruct its polarisation distribution. The scheme of the protocol is sketched in Fig.~\ref{fig:schemelight}. In this case, the light acts as the ancillary system ($D$) which initially is a squeezed state
with variance $\sigma^2$.

The resulting light mode distribution faithfully reproduces the work probability distribution provided that the ratio $\sigma/\kappa$ is sufficiently smaller than the energy differences. Finding the first moments of this distribution we obtain:
\begin{eqnarray}
\langle X\rangle &=& \kappa\langle W\rangle 
\\
\langle X^2\rangle &=&\sigma^2 +\kappa^2\langle W^2\rangle 
\end{eqnarray}
which clearly shows how to obtain the first moment of the work by analysing the fluctuations of the light mode quadrature. However it is clear from the expression of the second moment, that the initial zero-point fluctuations of the light mode will contribute with additional noise and may render difficult the estimate of $\langle W^2\rangle$ if the condition $\sigma/\kappa\ll 1$ is not satisfied. 
The situation is dramatically worse if one wants to reconstruct the Jarzynski equality since the noise originating from the light mode will contribute to all orders and exponentially increasing. Indeed:
\begin{equation}
\langle e^{-\beta X/\kappa} \rangle = \langle e^{-\beta W} \rangle\exp\left(\frac{\beta^2\sigma^2}{2\kappa^2}\right)
\end{equation}
showing that the Jarzynski equality is affected by an exponential correction that diverges with $\sigma/\kappa$ and in the limit of small temperatures \cite{DeChiaraNJP2015, talkner2016,solinas2017measurement}.

\subsection{Experimental realization of a quantum work meter with cold atoms}

The POVM proposed in Ref.~\cite{roncaglia2014work} was recently realised experimentally with cold atoms in Ref.~\cite{Cerisola2017}. In this experimental setup, a Bose-Einstein condensate (BEC) of ${}^{87}$Rb atoms with two internal Zeeman sublevels $\ket 1\equiv \ket{F=2,m_F=1}$ and  $\ket 2\equiv \ket{F=2,m_F=2}$ is trapped by an atom chip. The BEC is then released from the trap and, during the free fall, a spatially inhomogeneous magnetic field created by the atom chip is capable of realising a Stern-Gerlach type of interaction \cite{machluf2013coherent}. Such a process entangles the spin degree of freedom (the system of which the work injected will be measured) with the motional degree of freedom of the atoms.  This is exactly the type of interaction of Eq.~\eqref{eq:uent} that was proposed in Ref.~\cite{roncaglia2014work,DeChiaraNJP2015}.

Thus, the key element of the experiment is
the atom chip, which efficiently entangles the internal and motional
degrees of freedom of an atom through Stern-Gerlach type magnetic gradient pulses.
These pulses are generated using a $3$-current-carrying-wire setup on the chip surface.
A gradient pulse along the vertical direction $z$ direction with amplitude $B'$ and
duration $\tau$, induces a momentum kick
$m_F\, \delta p$ on an atom in the $m_F$ state
($\delta p\sim\mu_B g_F B'\tau$, where $\mu_B$
and $g_F$ are, respectively, the Bohr magneton and the Land\'e factor\,\cite{machluf2013coherent}).
The evolution of the state of the atom induced by such
a pulse is described by the unitary operator
\begin{equation}
U_p=e^{i \,\delta p\, z_D  \sigma_S},
\end{equation}
where the operator $ z_D$ is the generator of translations in the momentum degree of freedom, and
 ${\sigma_S=\ket{1}\bra{1}+2\ket{2}\bra{2}}$. If the Hamiltonian of the 
atom is proportional to the operator $\sigma_S$ this operation effectively implements
the evolution of Eq.~\eqref{eq:uent}. In this case the motional degree of freedom of the atoms
is the ancillary system or detector. The momentum kicks induced by both
pulses are controlled in the experiment, and it is also possible to apply an RF field before the entangling operation.
Thus, with these tools it is possible to simulate an arbitrary system
with initial and final Hamiltonians $H_S(0)$ and $H_S(\mathcal T)$~\cite{roncaglia2014work}.
Finally, let us note that the first pulse $B'$ and the second pulse $\tilde B'$  are applied with opposite signs to ensure that
the sequence creates a coherent record of the value of work.

The initial state of the spins is a pure superposition of states $\ket 1$ and $\ket 2$ in such a way that the populations 
correspond to a thermal state with temperature $T$. Since in the experiment, one is interested in the realisation of the \tmp{} by a single POVM, 
one can adjust the value of $\delta p$ such that coherences do not affect the experiment (see Ref.~\cite{Cerisola2017}  and its 
Supplementary Material). The experiment proceeds as described before: $(i)$ the atoms are released from the atom chip and while falling, $(ii)$ internal and motional degrees of freedom are entangled with a magnetic gradient pulse, $(iii)$ a unitary driving $U_S(\mathcal T)$ of the spin state is applies through a RF field. 
$(iv)$ Another magnetic gradient field couples internal and motional degrees of freedom, $(v)$ after a free fall evolution an image of the atoms is taken. 

\begin{figure}[t]
\begin{center}
\includegraphics[width=\columnwidth]{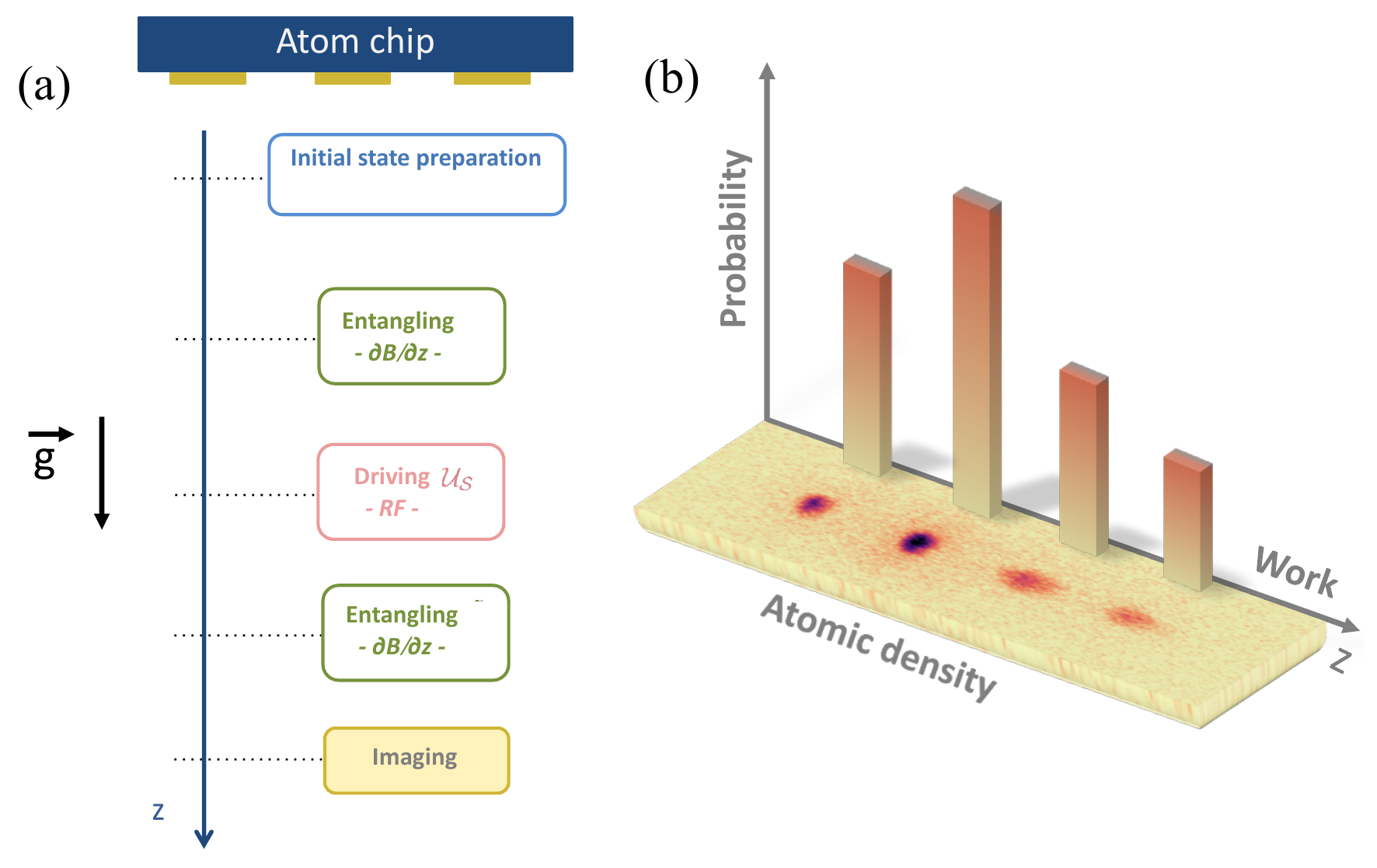}
\caption{(a) Physical operations that allow to measure work on the spin state of a BEC under RF driving (see text). The ancillary system consists of the 
motional degrees of freedom of the atoms.
(b) Probability distribution of work. The image shows the BEC split in 4 clouds at the end of a single run of the experiment. The $z$ axis measures the position of the cloud and can be put in correspondence with the work done on the spins. The vertical bars correspond to the number of atoms in each of the 4 clouds and represent the probability for the corresponding work values. Figure reproduced  from Ref.~\cite{Cerisola2017} under the CC BY licence.}
\label{fig:workBEC}
\end{center}
\end{figure}

The above sequence and an example of the atomic image are shown in Fig.~\ref{fig:workBEC}. The 4 peaks correspond to the 4 processes corresponding to all the transitions from the initial to the final states (similar to the Ramsey experiment in NMR). 
The position of the peak can be connected to the value of the work while the corresponding population is proportional to the probability for that value.
It is interesting to notice that in a single run of the experiment the full probability distribution is obtained.
The authors of Ref.~\cite{Cerisola2017} were also able to verify the Jarzynski equality for different initial temperatures of the sample and different processes $U_S(\mathcal T)$. 

\section{Conclusions}
\label{sec:conclusions}

In summary, we have considered protocols that allow the reconstruction of the probability distribution of work done on a quantum system using an auxiliary quantum detector.  We described two paradigmatic schemes giving directly access to the characteristic function and the distribution of work, and their 
recent physical implementations using different setups.
The necessary ingredient for these schemes is a system-detector interaction that couples the system energy operator with an observable of the detector. 
Then, we described a framework that provides a unified view of these ideas.
In this framework fall most of the proposals and experimental measurements of heat and work that appeared in the literature in recent years.
The resulting work statistics and the degree of perturbation induced in the system dynamics depend on the system-detector coupling and 
on the measurement done on the detector. Therefore, the protocol used to measure the work statistics must be chosen keeping in mind which quantum features we want to preserve in the dynamics. 
 
We believe that there are many unanswered questions on the role of quantum energy coherences in thermodynamic processes. 
The schemes described in this chapter are important tools for such future investigations.

\acknowledgments

We thank R. Fazio, J. Goold, L. Mazzola, K. Modi, M. Paternostro and J. P. Paz for illuminating discussions.
P.S. has received funding from the European Union FP7/2007-2013 under REA Grant Agreement No. 630925, COHEAT, and from MIUR-FIRB2013, Project Coca (Grant No. RBFR1379UX). FC and AJR acknowledge financial support from ANPCyT (PICT 2013-0621 and PICT 2014-3711), CONICET
and UBACyT.

\bibliography{Quantum_thermo_biblio}

\end{document}